\documentclass{article}
\usepackage{PRIMEarxiv}
\usepackage[hyphens]{url}
\usepackage{amsmath}
\usepackage{graphicx}
\usepackage{pythonhighlight}
\usepackage{hyperref}

\title{
    OkadaTorch: A Differentiable Programming of Okada Model to Calculate Displacements and Strains from Fault Parameters
}

\author{
    Masayoshi Someya \\ \\
    Earthquake Reseach Institute,\\
    The University of Tokyo \\
    \texttt{someya@eri.u-tokyo.ac.jp} \\
    \And
    Taisuke Yamada \\ \\
    Research Center for Prediction of \\
    Earthquakes and Volcanic Eruptions, \\
    Tohoku University
    \AND
    Tomohisa Okazaki \\ \\
    RIKEN Center for Advanced Intelligence Project \\
}

\begin{document}
\maketitle

\begin{abstract}
    
The Okada model is a widely used analytical solution for displacements and strains caused by a point or rectangular dislocation source in a 3D elastic half-space.
We present \pyth{OkadaTorch}, a PyTorch implementation of the Okada model, where the entire code is differentiable; gradients with respect to input can be easily computed using automatic differentiation (AD).
Our work consists of two components: a direct translation of the original Okada model into PyTorch, and a convenient wrapper interface for efficiently computing gradients and Hessians with respect to either observation station coordinates or fault parameters.
This differentiable framework is well suited for fault parameter inversion, including gradient-based optimization, Bayesian inference, and integration with scientific machine learning (SciML) models.
Our code is available here: \url{https://github.com/msomeya1/OkadaTorch}

\end{abstract}

\section{Introduction}

The Okada model \cite{okada1985surface, okada1992internal} provides an analytical solution for displacements and strains (spatial derivatives of displacement) caused by a point or rectangular dislocation source in a 3D elastic half-space. It has become a standard tool in seismology and geodesy for modeling coseismic deformation, and has been widely used to estimate fault slip distributions from GNSS/InSAR data \cite{ohta2012quasi, nishimura2013detection, ozawa2016crustal, kawamoto2017regard, bagnardi2018inversion} and tsunami data \cite{satake1994mechanism, tanioka1996fault, fujii2011tsunami}.

In addition to the original FORTRAN implementation \cite{DC3D_NIED}, various implementations or wrappers of Okada model exist, including Fortran \cite{DC3D.f90_Miyashita}, MATLAB \cite{Okadasolution_Dutykh, Okada_Beauducel, okada_wrapper}, Python \cite{okada_wrapper, okada4py, clawpack_okada, OkadaPy}, and Julia \cite{Tsunami.jl}.
These implementations are well suited for forward modeling and facilitate integration with models written in each programming language.

However, inverse modeling of crustal deformation sometimes requires the computation of model gradients with respect to input parameters \cite{matsu1977inversion, matsu1987maximum}.
One traditional approach is to derive the analytical derivatives of the forward model, either by hand \cite{matsu1977inversion} or with the aid of computer algebra system \cite{pires2001tsunami}, and then implement these expressions in programs. 
This approach is known as \textit{symbolic differentiation}.
An alternative is \textit{automatic differentiation} (AD);
it treats a numerical program as a composition of differentiable primitive operations and algorithmically applies the chain rule to compute exact derivatives of the program output with respect to its inputs.
Since AD can provide accurate gradients efficiently, it has become a core technology in deep learning frameworks. In recent years, a variety of AD libraries have been developed such as PyTorch \cite{paszke2019pytorch, ansel2024pytorch}, TensorFlow \cite{abadi2016tensorflow}, and JAX \cite{frostig2018compiling, jax2018github}.

AD has been used as a tool for inverse analysis in geophysics, especially applied to inversion of seismic source and subsurface structures \cite{sambridge2007automatic, zhu2021general, rasht2022physics, yang2023rapid}. However, to our knowledge, there has been no attempt to apply AD to Okada model.
In this work, we present a PyTorch implementation of the Okada model that enables efficient computation of derivatives of displacement and strain.
Our implementation consists of two components:
\begin{itemize}
    \item A direct translation of the original Okada's subroutines, and
    \item A user-friendly wrapper interface, \pyth{OkadaWrapper}, that allows easy computation of gradients and Hessians.
\end{itemize}
This model serves as an essential component of differentiable programming in geophysical modeling, opens up new possibilities for advanced applications, including gradient-based inversion, sensitivity analysis, Bayesian inference, and integration with scientific machine learning (SciML) models.

\section{Direct Translation of the Original Okada's Subroutines}

The core of our implementation is a direct translation of the original FORTRAN subroutines into PyTorch functions:

\begin{itemize}
    \item \pyth{SPOINT} \cite{okada1985surface}: Calculate displacements and strains at the surface ($z=0$) created by a point source.
    \item \pyth{SRECTF} \cite{okada1985surface}: Calculate displacements and strains at the surface ($z=0$) created by a rectangular fault.
    \item \pyth{DC3D0} \cite{okada1992internal}: Same as \pyth{SPOINT}, but under the surface ($z\leq0$).
    \item \pyth{DC3D} \cite{okada1992internal}: Same as \pyth{SRECTF}, but under the surface ($z\leq0$).
\end{itemize}

These functions are implemented using PyTorch tensor operations, enabling vectorized computation over multiple stations and GPU acceleration.

The interface of each function closely follows that of the original FORTRAN version. For details of the core algorithm and usage, we refer readers to the original publications \cite{okada1985surface, okada1992internal, DC3D_NIED}. However, we introduce two additional keyword arguments in our implementation that are not present in the original code: \pyth{compute_strain} and \pyth{is_degree}.

\begin{itemize}
    \item \pyth{compute_strain}: A boolean flag indicating whether to compute strain components in addition to displacements.
    The original FORTRAN subroutines always return both, but in some cases, displacement alone is sufficient.
    When \pyth{compute_strain} is \pyth{False}, intermediate variables that are only used to compute strain are not assigned, thus reducing computational cost.
    \item \pyth{is_degree}: A boolean flag indicating whether angular parameters (strike, dip, rake) are specified in degrees (\pyth{True}, default) or radians (\pyth{False}).
\end{itemize}

Below is a simple example that demonstrates how to compute displacements using the \pyth{DC3D} function without strain:

\begin{python}
import numpy as np
import torch
from OkadaTorch import DC3D

ALPHA = 2.0/3.0
x = np.linspace(-1, 1, 101)
y = np.linspace(-1, 1, 101)
z = np.linspace(-1, 0, 51)
X, Y, Z = np.meshgrid(x, y, z)
X = torch.from_numpy(X)
Y = torch.from_numpy(Y)
Z = torch.from_numpy(Z)

DEPTH = 2.0
DIP = torch.tensor(45.0)
AL1, AL2 = -0.2, 0.2
AW1, AW2 = -0.1, 0.1

DISL1, DISL2, DISL3 = 4.0, 3.0, 0.0

out, IRET = DC3D(ALPHA, X, Y, Z, DEPTH, DIP, 
                 AL1, AL2, AW1, AW2, 
                 DISL1, DISL2, DISL3, 
                 compute_strain=False, is_degree=True)
\end{python}

In this case, \pyth{out} is a list of three tensors $\left[u_x, u_y, u_z\right]$, corresponding to the displacement components along the $x$, $y$, and $z$ axes.
\pyth{IRET} is an integer flag inherited from the original FORTRAN implementation, indicating whether the computation was successful.
Both the displacements and \pyth{IRET} have the same shape as the input coordinate tensors \pyth{X}, \pyth{Y}, and \pyth{Z}.
In the original FORTRAN implementation, the subroutines are designed to operate on scalar inputs, and to compute displacements at multiple stations, users need to explicitly loop over all coordinates and call the subroutine repeatedly. In contrast, our PyTorch implementation supports fully vectorized tensor inputs, allowing all stations to be processed in a single function call without any explicit looping. 
This not only simplifies the code but also improves computational efficiency, especially on GPUs.

Examples demonstrating the use of other subroutines, such as \pyth{SPOINT}, \pyth{SRECTF}, and \pyth{DC3D0}, are available in the \pyth{OkadaTorch} GitHub repository. In practice, users may be interested in computing displacements either at the surface or under the surface, depending on the application. To simplify such use cases, we provide a unified wrapper interface that internally selects the appropriate subroutine based on the input configuration. This interface, along with its support for automatic differentiation, is described in the following section.

\section{The \pyth{OkadaWrapper} Class}

\begin{figure*}
    \centering
    \includegraphics[width=\textwidth]{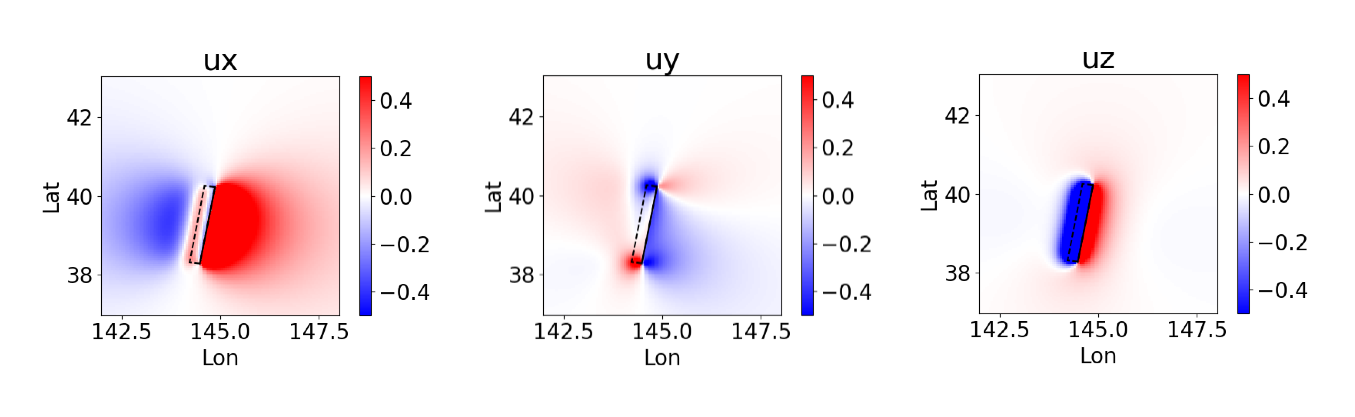}
    \caption{
        Three displacement components calculated by \pyth{compute} method (units: m).
        The dashed rectangles represent the surface projections of the faults,
        and the upper edges along the strike directions are shown as solid lines.
    }
    \label{fig:ui}
\end{figure*}

\begin{figure*}
    \centering
    \includegraphics[width=\textwidth]{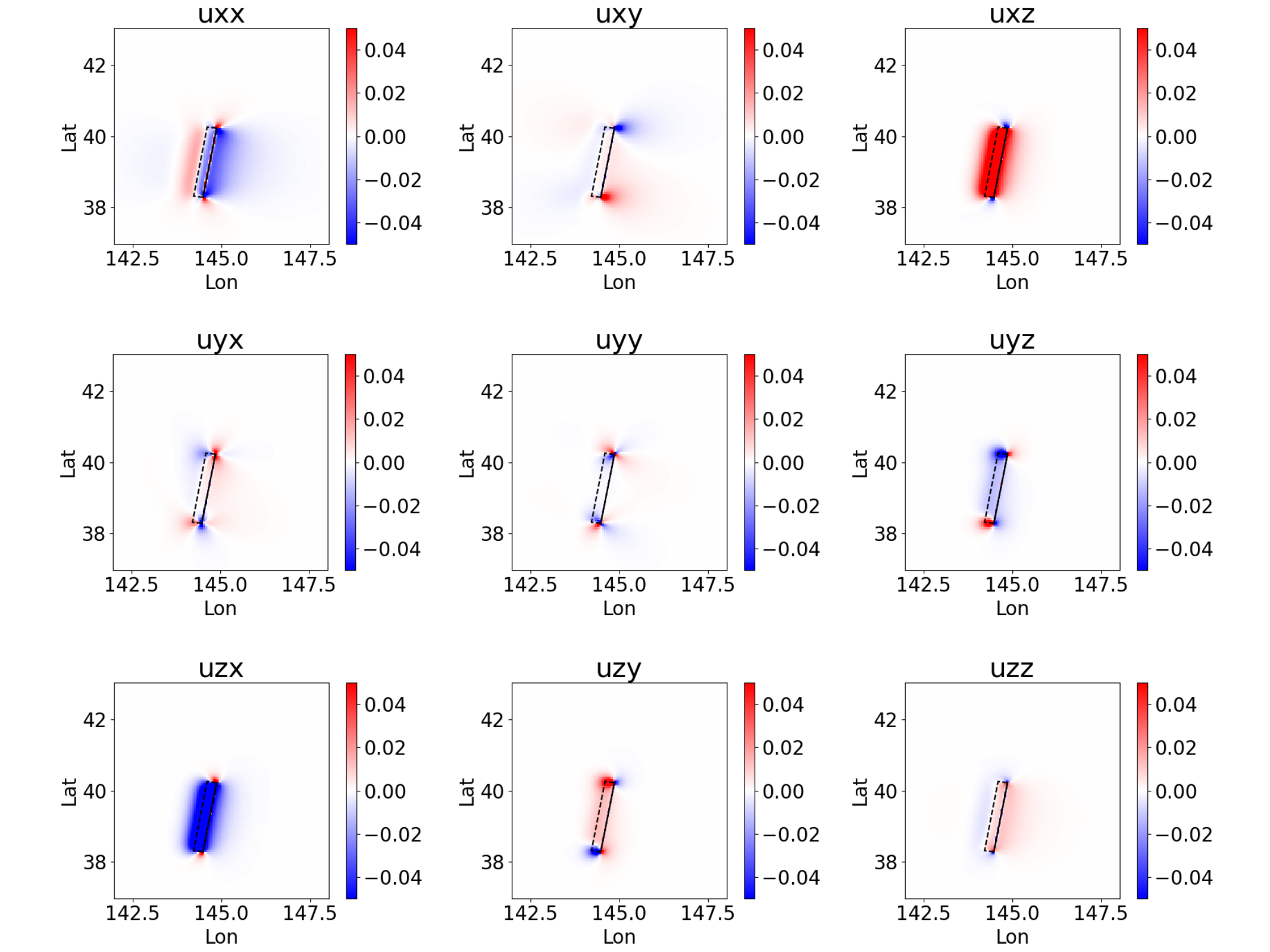}
    \caption{
        Nine strain components calculated by \pyth{compute} method
        (units: m/km, displacement [m] differentiated with respect to distance [km]).
    }
    \label{fig:uij}
\end{figure*}

To simplify the use of functions and to easily calculate gradients and Hessians, we provide a high-level wrapper class, \pyth{OkadaWrapper}. This class abstracts the low-level subroutines (\pyth{SPOINT}, \pyth{SRECTF}, \pyth{DC3D0}, \pyth{DC3D}) and offers a unified interface for forward modeling and derivative computation.

This class supports three main methods: \pyth{compute}, \pyth{gradient}, and \pyth{hessian}. Each method accepts the same basic inputs: observation coordinates (\pyth{coords}) and fault parameters (\pyth{params}), with additional arguments specific to each method. In the following subsections, we describe each method in turn, along with examples.

\subsection{Forward Modeling with \pyth{compute} method}

The \pyth{compute} method performs the forward calculation; given the fault parameters, the displacements and/or strains are calculated.

Required arguments are as follows.
\begin{itemize}
    \item \pyth{coords}: A Python dictionary containing observation station coordinates.
    Allowed keys are \pyth{"x"}, \pyth{"y"}, and optionally \pyth{"z"} (all PyTorch tensors of the same shape).
    The coordinate system is right-handed: $x$ (east), $y$ (north), $z$ (upward). Note that $z$ must be negative for subsurface observation points.

    \item \pyth{params}: A dictionary specifying the fault parameters.
    Required keys include:

    \begin{itemize}
        \item \pyth{"x_fault"}, \pyth{"y_fault"}, \pyth{"depth"}: $x, y$ coordinates and depth of the source (depth is positive).
        In the case of a point source, these values represent the location of that point. In the case of a rectangular fault, the flag \pyth{fault_origin} specifies which point these values represent. If \pyth{fault_origin} is \pyth{"topleft"}, then \pyth{"x_fault"}, \pyth{"y_fault"} and \pyth{"depth"} represent the coordinates of the top left corner of the rectangle. If \pyth{fault_origin} is \pyth{"center"}, then \pyth{"x_fault"}, \pyth{"y_fault"} and \pyth{"depth"} represent the coordinates of the rectangle's center.
        \item \pyth{"strike"}, \pyth{"dip"}, \pyth{"rake"}: fault orientation.
        \item \pyth{"slip"}: slip amount (for rectangular faults) or potency (for point sources).
    \end{itemize}
    Optional keys: \pyth{"length"} and \pyth{"width"} for rectangular faults.
\end{itemize}

Other optional arguments are as follows.

\begin{itemize}
    \item \pyth{compute_strain}: whether to compute strain components in addition to displacements. Default \pyth{True}.
    \item \pyth{is_degree}: whether strike/dip/rake are in degrees. Default \pyth{True}.
    \item \pyth{fault_origin}: which point the fault location parameters (\pyth{"x_fault"}, \pyth{"y_fault"} and \pyth{"depth"}) refer to. Either \pyth{"topleft"} (default) or \pyth{"center"} can be specified.
    \item \pyth{nu}: Poisson's ratio of the medium. Default \pyth{0.25}.
\end{itemize}

A typical usage is as follows:

\begin{python}
from OkadaTorch import OkadaWrapper

coords = {
    "x": x, # torch.Tensor representing x-coordinate of the station
    "y": y  # torch.Tensor representing y-coordinate of the station
}
params = { # All values are torch.Tensor (each is a scalar)
    "x_fault": x_fault,
    "y_fault": y_fault,
    "depth": depth,
    "length": length,
    "width": width,
    "strike": strike,
    "dip": dip,
    "rake": rake,
    "slip": slip
}

okada = OkadaWrapper()
out = okada.compute(coords, params)
\end{python}

If \pyth{compute_strain} is \pyth{True}, \pyth{out} is a list of 12 tensors: 3 displacement components and 9 strain components:
\begin{equation}
    \left[u_x, u_y, u_z, \frac{\partial u_x}{\partial x}, \ldots , \frac{\partial u_z}{\partial z}\right]
\end{equation}
If \pyth{False}, only 3 displacement components are returned:
\begin{equation}
    \left[u_x, u_y, u_z\right]
\end{equation}

Figures \ref{fig:ui} and \ref{fig:uij} present a demonstration of the \pyth{okada.compute} method applied to a rectangular fault model. The fault parameters are taken from the model 10 of Table S1 in \cite{baba2021frequency}.
Figure \ref{fig:ui} illustrates the surface displacement field, while Figure \ref{fig:uij} shows the corresponding strain distribution computed at the surface.
The result confirms that the wrapper produces physically consistent deformation patterns.

\subsection{First-Order Derivatives with \pyth{gradient} method}

The \pyth{gradient} method computes the derivative of the model output with respect to a single input variable.
PyTorch's function \pyth{jacfwd} is used internally.

\begin{itemize}
    \item \pyth{arg}: The variable to differentiate with respect to.
    This must be one of the keys in either \pyth{coords} (e.g., \pyth{"x"}, \pyth{"y"}, \pyth{"z"}) or \pyth{params} (e.g., \pyth{"depth"}, \pyth{"strike"}, etc.).
\end{itemize}

If \pyth{arg} is one of the fault parameters, the resulting first-order derivatives represent the sensitivity of the output with respect to that parameter. These sensitivities can be directly used for gradient-based first-order optimization methods, as well as for local sensitivity analysis and parameter studies.

Other arguments (\pyth{coords}, \pyth{params}, \pyth{compute_strain}, \pyth{is_degree}, \pyth{fault_origin} and \pyth{nu}) are the same as in the \pyth{compute} method.

A typical usage is as follows:

\begin{python}
out = okada.gradient(coords, params, arg="x")
\end{python}

\begin{python}
out = okada.gradient(coords, params, arg="depth")
\end{python}

If \pyth{compute_strain} is \pyth{True}, \pyth{out} is a list of 3 displacement components and 9 strain components differentiated by \pyth{arg}:
\begin{equation}
    \left[\frac{\partial u_x}{\partial\text{(arg)}}, \frac{\partial u_y}{\partial\text{(arg)}}, \frac{\partial u_z}{\partial\text{(arg)}}, \frac{\partial}{\partial\text{(arg)}}\left(\frac{\partial u_x}{\partial x}\right), \ldots, \frac{\partial}{\partial\text{(arg)}}\left(\frac{\partial u_z}{\partial z}\right)\right]
\end{equation}
If \pyth{False}, \pyth{out} is a list of 3 displacement components differentiated by \pyth{arg}:
\begin{equation}
    \left[\frac{\partial u_x}{\partial\text{(arg)}}, \frac{\partial u_y}{\partial\text{(arg)}}, \frac{\partial u_z}{\partial\text{(arg)}}\right]
\end{equation}

\begin{figure*}
    \centering
    \includegraphics[width=\textwidth]{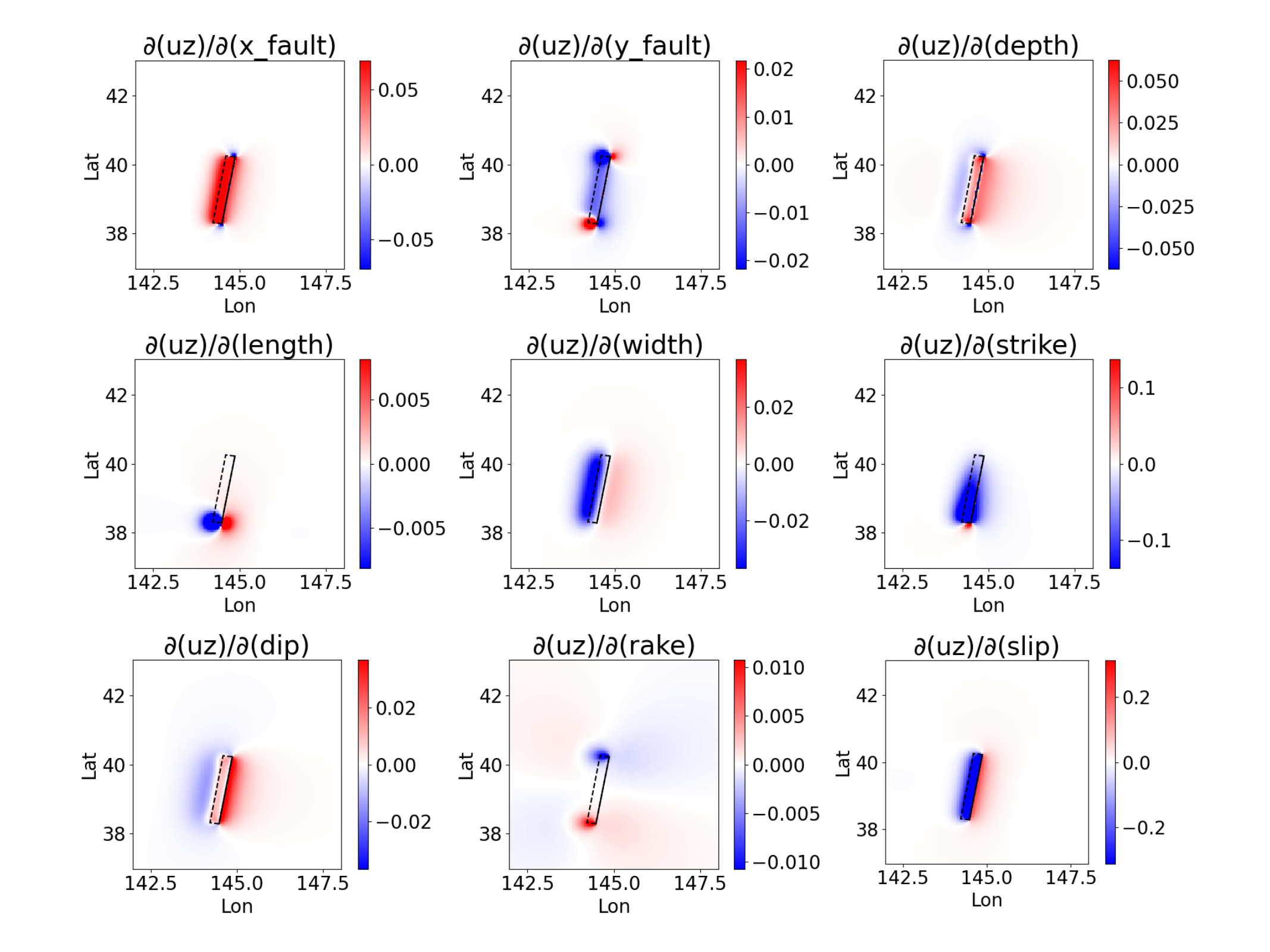}
    \caption{
        First-order derivatives of the vertical displacement \pyth{uz} with respect to fault parameters, calculated by \pyth{gradient} method.
    }
    \label{fig:grad}
\end{figure*}

Figure \ref{fig:grad} shows the first-order derivatives of the vertical displacement $u_z$ with respect to nine fault parameters (same parameters as in Figures \ref{fig:ui} and \ref{fig:uij}).
As expected, the derivative with respect to slip --— being linearly related to the output --— exhibits the same spatial pattern as $u_z$ itself. 
In contrast, derivatives with respect to the other nonlinear parameters display spatially distinct and interpretable structures. 
Such visualizations may also be useful for educational purposes, for example, to help students new to seismology understand how each fault parameter influences observed surface deformation.

\begin{figure*}
    \centering
    \includegraphics[width=\textwidth]{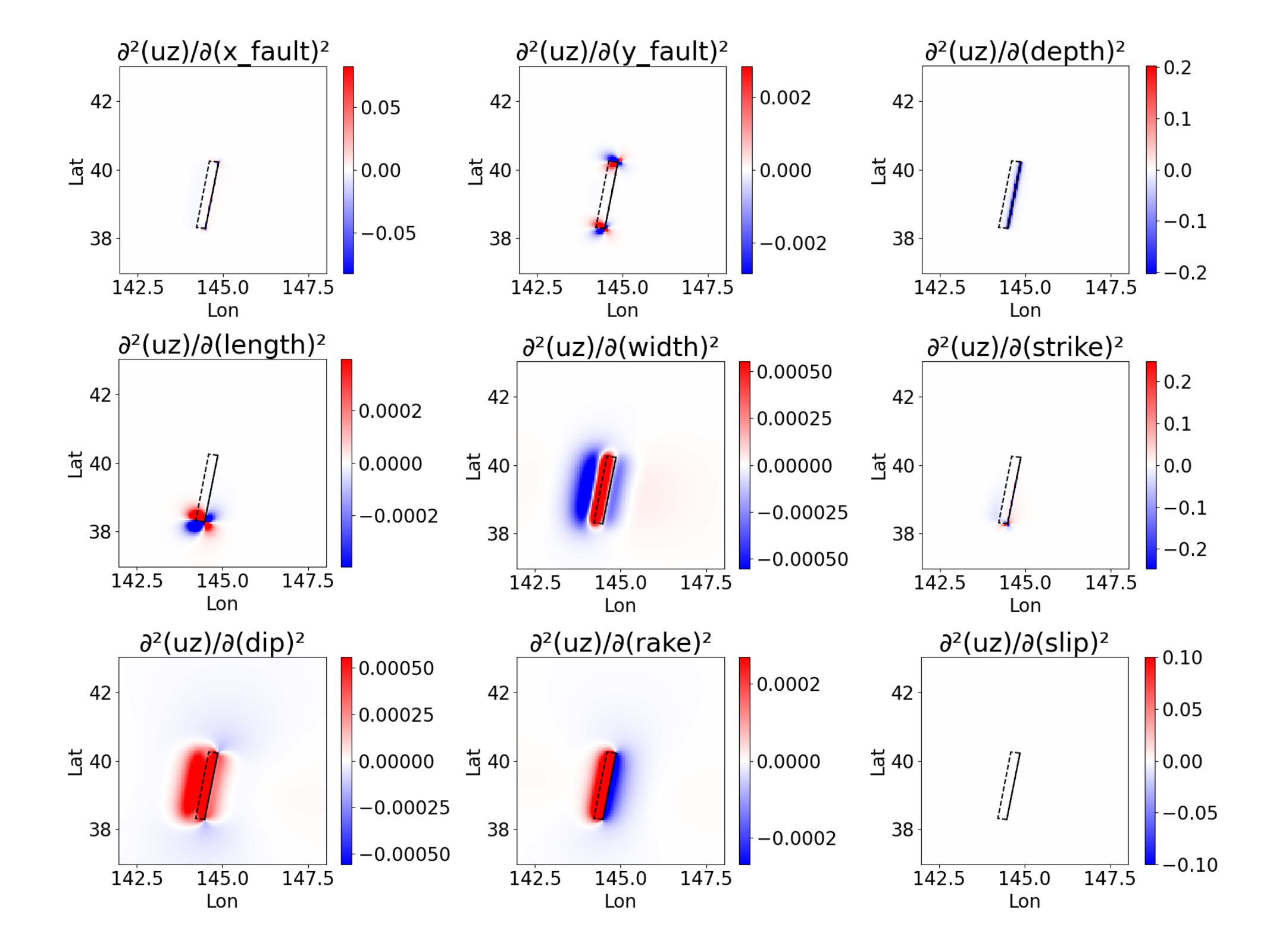}
    \caption{
        Diagonal Hessian of vertical displacement \pyth{uz} with respect to fault parameters, computed by \pyth{hessian} method.
    }
    \label{fig:hessian}
\end{figure*}

\subsection{Second-Order Derivatives with \pyth{hessian} method}

The \pyth{hessian} method computes the second-order derivatives of the model output with respect to two variables.
PyTorch's function \pyth{jacfwd} is used internally.

\begin{itemize}
    \item \pyth{arg1}, \pyth{arg2}: The variables to differentiate with respect to.
    These must both be from the same category (either both from \pyth{coords} or both from \pyth{params}).
    For instance, \pyth{("x", "y")} or \pyth{("depth", "rake")} are valid pairs, but \pyth{("x", "slip")} is not.
\end{itemize}

If both \pyth{arg1} and \pyth{arg2} are fault parameters, the resulting second-order derivatives quantify the local curvature of the forward model output with respect to those parameters. 
These second derivatives are useful for second-order optimization, uncertainty quantification, and Laplace approximation methods in Bayesian inference, where knowledge of the Hessian around the optimum is essential.

Other arguments (\pyth{coords}, \pyth{params}, \pyth{compute_strain}, \pyth{is_degree}, \pyth{fault_origin} and \pyth{nu}) are the same as in the \pyth{compute} method.

A typical usage is as follows:

\begin{python}
out = okada.hessian(coords, params, arg1="x", arg2="y")
\end{python}

\begin{python}
out = okada.hessian(coords, params, arg1="depth", arg2="depth")
\end{python}

If \pyth{compute_strain} is \pyth{True}, \pyth{out} is a list of 3 displacement components and 9 strain components differentiated by \pyth{arg1} and \pyth{arg2}:
\begin{equation}
    \left[\frac{\partial^2 u_x}{\partial\text{(arg1)}\partial\text{(arg2)}}, \frac{\partial^2 u_y}{\partial\text{(arg1)}\partial\text{(arg2)}}, \frac{\partial^2 u_z}{\partial\text{(arg1)}\partial\text{(arg2)}}, \frac{\partial^2}{\partial\text{(arg1)}\partial\text{(arg2)}}\left(\frac{\partial u_x}{\partial x}\right), \ldots, \frac{\partial^2}{\partial\text{(arg1)}\partial\text{(arg2)}}\left(\frac{\partial u_z}{\partial z}\right)\right]
\end{equation}
If \pyth{False}, \pyth{out} is a list of 3 displacement components differentiated by \pyth{arg1} and \pyth{arg2}:
\begin{equation}
    \left[\frac{\partial^2 u_x}{\partial\text{(arg1)}\partial\text{(arg2)}}, \frac{\partial^2 u_y}{\partial\text{(arg1)}\partial\text{(arg2)}}, \frac{\partial^2 u_z}{\partial\text{(arg1)}\partial\text{(arg2)}}\right]
\end{equation}

Figure \ref{fig:hessian} shows diagonal entries of the Hessian of $u_z$, corresponding to cases where \pyth{arg1} = \pyth{arg2}.
As expected, the second derivative with respect to slip is zero, since $u_z$ depends linearly on slip. 
In contrast, the second derivatives with respect to nonlinear parameters exhibit distinct spatial patterns, reflecting the complex ways in which each parameter influences the surface deformation.

\subsection{Example Application: Parameter Estimation by Loss Function Optimization}

\begin{figure*}
    \centering
    \includegraphics[width=\textwidth]{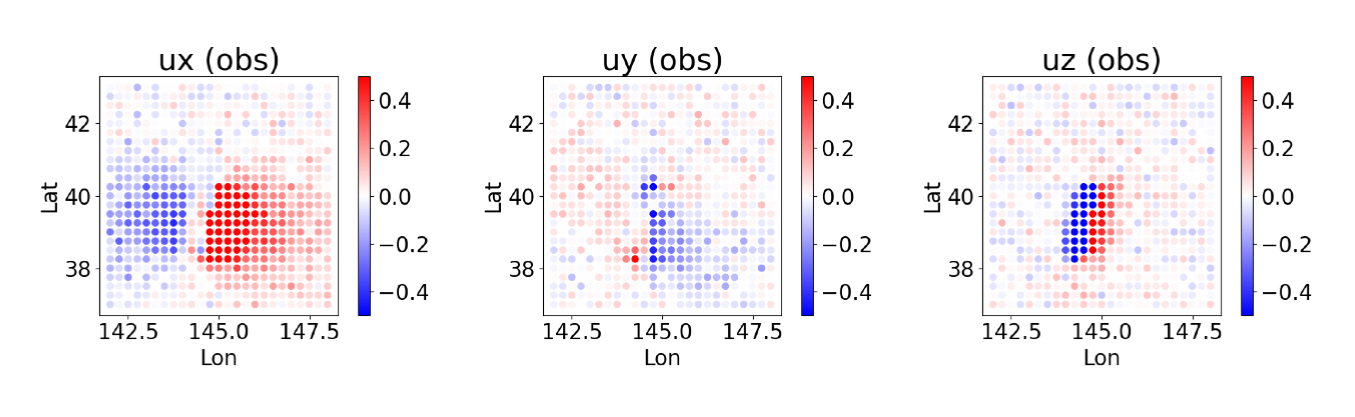}
    \caption{
    Synthetic observation data generated by adding random noise to the forward modeling results. 
    The grid interval is 0.25 degrees.
    }
    \label{fig:uobs}
\end{figure*}

\begin{figure*}
    \centering
    \includegraphics[width=\textwidth]{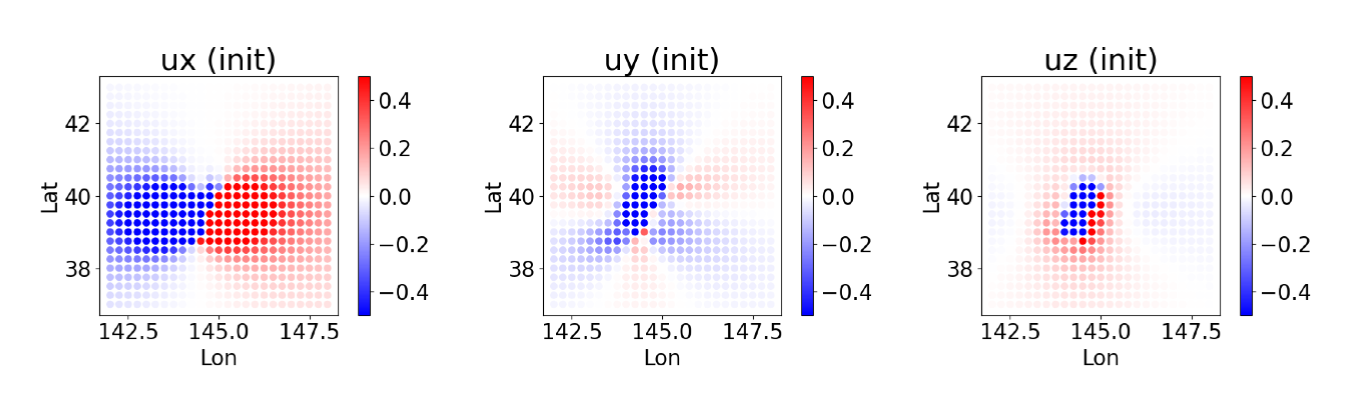}
    \caption{
        Displacement components computed using the initial fault parameters at the start of the optimization.
    }
    \label{fig:uinit}
\end{figure*}

\begin{figure*}
    \centering
    \includegraphics[width=\textwidth]{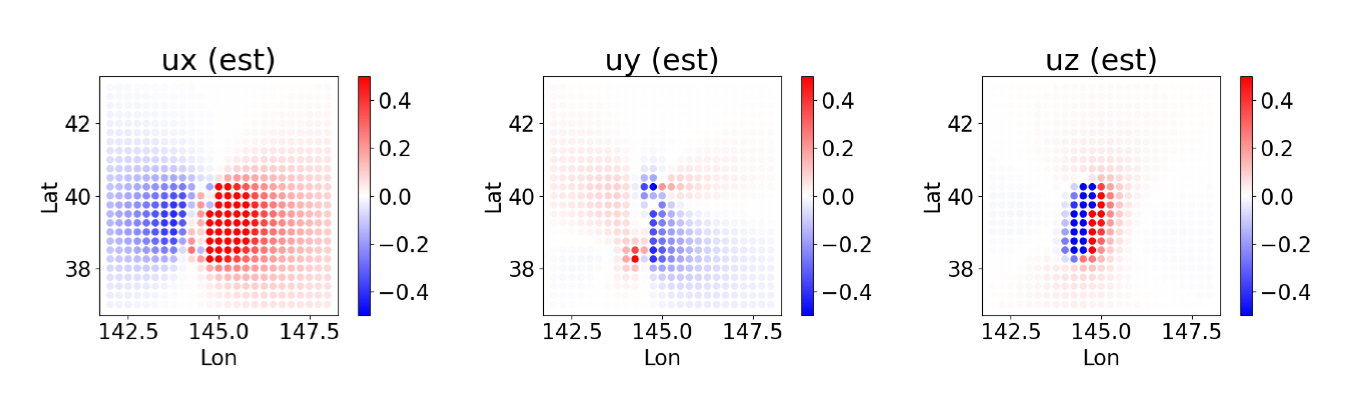}
    \caption{
        Estimated displacement components computed using the optimized fault parameters.
    }
    \label{fig:uest}
\end{figure*}

\begin{figure*}
    \centering
    \includegraphics[width=0.7\textwidth]{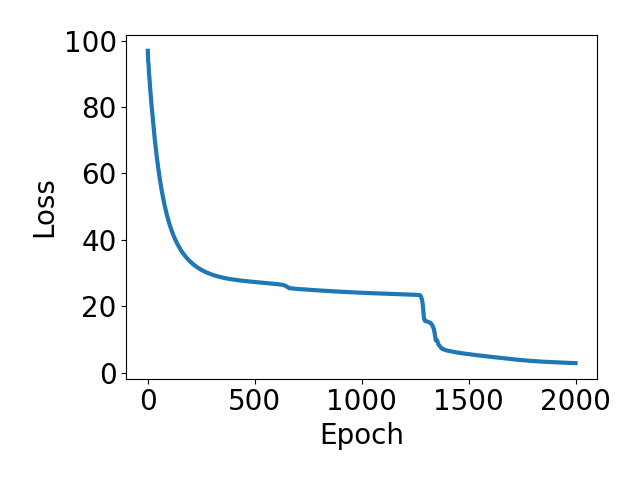}
    \caption{
        Evolution of the loss function over 2000 optimization epochs.
    }
    \label{fig:loss}
\end{figure*}

\begin{table}
    \centering
    \caption{
        True values of fault parameters used in numerical experiments \cite{baba2021frequency}, initial guess before optimization, and final values after optimization.
        Units are km from \pyth{x_fault} to \pyth{width}, degrees for \pyth{strike}, \pyth{dip}, and \pyth{rake}, and m for \pyth{slip}.
    }
    \label{tab:params}
    \begin{tabular}{lccccccccc}
        \hline
        & \texttt{x\_fault} & \texttt{y\_fault} & \texttt{depth} & \texttt{length} & \texttt{width} & \texttt{strike} & \texttt{dip} & \texttt{rake} & \texttt{slip} \\
        \hline 
        True & $-11.25$ & $24.70$ & $0.100$ & $218.0$ & $46.00$ & $189.0$ & $60.00$ & $270.0$ & $\phantom{0}5.62$ \\ 
        Initial & $\phantom{-1}0.00$ & $10.00$ & $1.000$ & $150.0$ & $60.00$ & $200.0$ & $45.00$ & $300.0$ & $10.00$ \\
        Final  & $-12.91$ & $23.06$ & $0.604$ & $210.4$ & $41.67$ & $189.1$ & $59.81$ & $269.6$ & $\phantom{0}6.14$ \\ 
        \hline 
    \end{tabular}
\end{table}

In this subsection, we demonstrate how fault parameters can be estimated from observed displacement data using gradient-based optimization.
Rather than using real geodetic observations, we generate synthetic data by adding random noise to surface displacements obtained from forward modeling. 

We begin by computing displacements using the same fault parameters as in Figure \ref{fig:ui}. 
Gaussian random noise with an amplitude of 0.05 m is then added to each component $(u_x, u_y, u_z)$ (Figure \ref{fig:uobs}).
Given these synthetic observations, we estimate the fault parameters by minimizing the misfit between observed and calculated displacements.
The optimization is performed using the Adam optimizer over 2000 epochs.
The core part of the code is excerpted as follows:
\begin{python}
import torch
from OkadaTorch import OkadaWrapper
okada = OkadaWrapper()

params = { # initialization
    "x_fault": torch.tensor(0.0, requires_grad=True),  
    "y_fault": torch.tensor(10.0, requires_grad=True),
    "depth": torch.tensor(1.0, requires_grad=True),
    "length": torch.tensor(150.0, requires_grad=True),
    "width": torch.tensor(60.0, requires_grad=True),
    "strike": torch.tensor(200.0, requires_grad=True),
    "dip": torch.tensor(45.0, requires_grad=True),
    "rake": torch.tensor(300.0, requires_grad=True),
    "slip": torch.tensor(10.0, requires_grad=True)
}
optimizer = torch.optim.Adam(
    [p for p in params.values() if p.requires_grad],
)
for iter in range(2000):
    optimizer.zero_grad()
    ux, uy, uz = okada.compute(coords, params, 
                               compute_strain=False, is_degree=True, 
                               fault_origin="topleft")
    loss = 0.5 * ((ux - ux_obs) ** 2 + 
                  (uy - uy_obs) ** 2 + 
                  (uz - uz_obs) ** 2).sum()
    loss.backward()
    optimizer.step()
\end{python}

Note that we do not explicitly call the \pyth{gradient} method here. 
Since the entire model is written in PyTorch, gradients can be automatically computed by defining a loss function and calling \pyth{loss.backward()}.
This allows for seamless integration into PyTorch’s optimization pipeline.

Table \ref{tab:params} (middle row) and Figure \ref{fig:uinit} show the initial fault parameter values and the corresponding displacement components.
Table \ref{tab:params} (bottom row) and Figure \ref{fig:uest} show the final optimized parameters and the displacement components they produce.
The evolution of the loss function is also shown in Figure \ref{fig:loss}.

All estimated fault parameters are reasonably close to their true values, indicating successful convergence of the optimization. 
However, some trade-offs between parameters are apparent. 
For example, both \pyth{depth} and \pyth{slip} are overestimated, suggesting that similar surface displacements can be explained by different combinations of fault geometry and slip amplitude.
We note that for some parameters, particularly the fault location parameters \pyth{x_fault} and \pyth{y_fault}, the choice of initial values can have a strong impact on optimization. If the initial guess is too far from the true values, the solution may diverge.
In practical applications, it may be advisable to fix such sensitive parameters and optimize only the remaining ones.

The purpose of this example is not to propose a robust inversion framework, but rather to illustrate that gradient-based parameter optimization is straightforward when using a fully differentiable implementation.
Applying this technique to real data would require more advanced treatment.

\section{Conclusion and Outlook}

We have presented a PyTorch-based implementation of the Okada model for computing displacements and strains due to a point or rectangular dislocation source in a 3D elastic half-space. The implementation is differentiable, vectorized, and easily extensible.

The differentiability of the model opens up a wide range of potential applications. 
Gradient-based inversion of fault parameters can be performed by leveraging PyTorch's built-in optimizers such as Adam, eliminating the need to implement optimization routines from scratch. This enables efficient estimation of fault geometry and slip distribution from geodetic observations such as GNSS or InSAR. While the framework is broadly applicable, our experiments also highlight that certain parameters (e.g., fault location) can pose challenges for optimization due to sensitivity or non-uniqueness.

Sensitivity analysis and uncertainty quantification can be performed using first- and second-order derivatives (gradients and Hessians) with respect to fault parameters. The availability of exact gradients makes the model particularly suitable for gradient-informed Bayesian inference methods such as Hamiltonian Monte Carlo (HMC), which can provide probabilistic estimates of parameter uncertainty \cite{ohno2021real, ohno2022rapid, yamada2022comparison}.

The differentiable nature of the implementation also enables seamless integration with other PyTorch-based machine learning (ML) models, including physics-informed neural networks \cite{okazaki2022physics, okazaki2025physics}. For example, if a ML model is trained to predict fault slip distributions from geometric or frictional properties of faults, its output can be directly fed into our Okada implementation to compute resulting surface deformation. Similarly, the output of our \pyth{OkadaTorch} model can be combined with differentiable tsunami solvers or ML-based surrogate models. These connections enable the construction of end-to-end differentiable models, where model components ranging from fault mechanics to tsunami propagation are represented in a unified way.

\section*{Availability}

Our PyTorch implementation of Okada model is publicly available at
\url{https://github.com/msomeya1/OkadaTorch}.
The repository includes:
\begin{itemize}
    \item Complete source code for the core and wrapper modules
    \item Example scripts and notebooks
    \item Documentation and usage instructions
\end{itemize}
Programs published in this repository are different from the original programs published in the NIED website \cite{DC3D_NIED}.
The authors have obtained permission from NIED to publish these programs.

\section*{Acknowledgements}
We are grateful to Yoshimitsu Okada and the National Research Institute for Earth Science and Disaster Resilience (NIED) for kindly granting permission to publish the \pyth{OkadaTorch} code and this preprint.
We also thank Ryoichiro Agata, Daisuke Sato (Japan Agency for Marine-Earth Science and Technology), Masayuki Kano (Tohoku University), Rikuto Fukushima (Stanford University) for their valuable feedback during the development of the code, which greatly improved the usability of the implementation.
The first author is supported by a JSPS Research Fellowship for Young Scientists (DC1).

\section*{Author Contributions}

M.S. designed the study and implemented the code.
T.Y. and T.O. reviewed the code and contributed suggestions.
M.S. drafted the manuscript and prepared the figures.
T.Y. and T.O. provided feedbacks on the manuscript.
All authors have read and approved the final version of the manuscript.

\bibliographystyle{elsarticle-num}  
\bibliography{refs}

\end{document}